\documentclass[prl,aps,amssymb,twocolumn,showpacs,floatfix]{revtex4}
\usepackage{graphics}
\usepackage{epsfig} 
\usepackage{dcolumn}
\usepackage{amsmath}
\begin{document}
\newcommand\dd{{\operatorname{d}}}
\newcommand\sgn{{\operatorname{sgn}}}
\def\Eq#1{{Eq.~(\ref{#1})}}
\def\Ref#1{(\ref{#1})}
\newcommand\e{{\mathrm e}}
\newcommand\cum[1]{  {\Bigl< \!\! \Bigl< {#1} \Bigr>\!\!\Bigr>}}
\newcommand\vf{v_{_\text{F}}}
\newcommand\pf{p_{_\text{F}}}
\newcommand\ef{{\varepsilon} _{\text{\sc f}}}
\newcommand\zf{z_{_\text{F}}}
\newcommand\zfi[1]{{z_{_\text{F}}}_{#1}}
\newcommand\av[1]{\left<{#1}\right>}
\def\det{{\mathrm{det}}}
\def\Tr{{\mathrm{Tr}}}
\def\Li{{\mathrm{Li}}}
\def\tr{{\mathrm{tr}}}
\def\im{{\mathrm{Im}}}
\def\Texp{{\mathrm{Texp}\!\!\!\int}}
\def\antiTexp{{\mathrm{\tilde{T}exp}}\!\!\!\int}
\title{Full counting statistics of Luttinger liquid conductor}

\author{D. B. Gutman$^{1}$, Yuval Gefen$^2$, and A. D. Mirlin$^{3,4,5}$}
\affiliation{
\mbox{$^1$Department of Physics, Bar Ilan University, Ramat Gan 52900,
Israel }\\
\mbox{$^2$Dept. of Condensed Matter Physics, Weizmann Institute of
  Science, Rehovot 76100, Israel}\\
\mbox{$^3$Institut f\"ur Nanotechnologie, Karlsruhe Institute of Technology,
 76021 Karlsruhe, Germany}\\
\mbox{$^4$Institut f\"ur Theorie der kondensierten Materie,
Karlsruhe Institute of Technology, 76128 Karlsruhe, Germany}\\
\mbox{$^5$Petersburg Nuclear Physics Institute, 188300 St.~Petersburg, Russia}
}

\date{\today}

\begin{abstract}
Non-equilibrium bosonization technique is used to study current
fluctuations of interacting electrons in a single-channel quantum
wire representing a  Luttinger liquid (LL) conductor. An exact
expression for the full counting statistics of the  transmitted
charge is derived. It is given by Fredholm determinant of the
counting operator with a time dependent scattering phase. The result
has a form of counting statistics of
non-interacting particles with fractional charges, induced by
scattering off the boundaries between the LL wire and the
non-interacting leads.
\end{abstract}
 \maketitle Fluctuations are
among the most fundamental concepts arising in statistical physics.
In recent years,  non-equilibrium noise was measured in a variety of
electronic systems, such as quantum point
contacts\cite{noise_tunnel_junctions}, diffusive mesoscopic
conductors\cite{noise_mesoscopic_conductors}, and fractional quantum
Hall edges\cite{Reznikov}; see Ref.~\onlinecite{blanter-review} for
review. With the decrease  of a  sample size, characterization of
current fluctuations by the second moment only becomes insufficient.
This has triggered  recent interest in higher-order correlation
functions of  current statistics. The third cumulant of the noise
was measured in recent experiments \cite{S3}. A
more complete characterization of  current fluctuations is the full
counting statistics (FCS), introduced by  Levitov {\it et al.}
\cite{Levitov-noise}. In its full extent this fascinating
theoretical approach yields information about  all moments of the
number of electrons transferred (over a given time interval) through
a given terminal in a multi-terminal system, current
cross-correlations and entanglement, and large current fluctuations.

 For non-interacting systems, the FCS of
non-equilibrium fluctuations is well understood by now within
several complementary approaches, including the Fredholm determinant
formalism \cite{Levitov-noise,counting-statistics}, the
$\sigma$-model field-theoretical description \cite{sigma-model}, as
well as  the kinetic theory of fluctuations and related methods
\cite{kinetic-theory}.

Much less remains  known  concerning  fluctuations in interacting
systems. This problem is of particular interest at low dimensions, 
where  interaction affects the nature of the system at hand
in an essential way. This is, in particular, the case for FCS of
current through a quantum dot in the Coulomb blockade regime
\cite{coulomb-blockade}. The interaction also affects dramatically
the physics of one-dimensional (1D) fermionic systems where a
strongly correlated state---Luttinger liquid (LL)---is
formed\cite{giamarchi}. Experimental realizations of LL include
carbon nanotubes, semiconductor, metallic and polymer nanowires, as
well as quantum Hall edges. Recent experiments studied the shot
noise in carbon nanotubes \cite{exp-noise-nanotubes}.
Non-equilibrium physics of carbon nanotubes and quantum Hall edges
has been  explored through  tunneling spectroscopy
\cite{tunnel-spectroscopy} and Mach-Zehnder-iterferometry
\cite{MachZehnder} respectively. The LL description is also relevant
to interacting 1D bosonic systems, cf. e.g.   recent experiments on
ultracold atomic gases that probe statistical properties of
inteference contrast, thus obtaining information on the full
distribution of quantum noise \cite{cold_atoms}.

Previous theoretical work on current fluctuations in LL mainly
focussed on the second moment (shot noise) \cite{theory-noise-LL}.
The most intriguing observation was related to manifestations of
fractional charges in  shot noise. However, full understanding of
the nature of charge transfer processes (and, in particular, of
charge fractionalization) requires the analysis of the FCS. The
latter has been studied in a biased LL with an impurity
\cite{FCS-LL-impurity}. While an analytical solution via the
thermodynamic Bethe ansatz can be found \cite{FCS-LL-impurity}, it
is in general implicit and very cumbersome.

In the present  work we study the statistical properties of current
fluctuations in  non-equilibrium  LL conductor with arbitrary energy
distribution of right- and left-movers. Employing the recently
developed non-equilibrium bosonization technique \cite{GGM_2009}, we
find an exact solution of this problem.  We show that the full
distribution of the current noise  is closely  related to the
phenomenon of charge fractionalization. Specifically, we demonstrate
that the FCS of LL  reduces to the one of a non-interacting model
with fractional charges calculated below.

\begin{figure}
\includegraphics[width=1\columnwidth,angle=0]{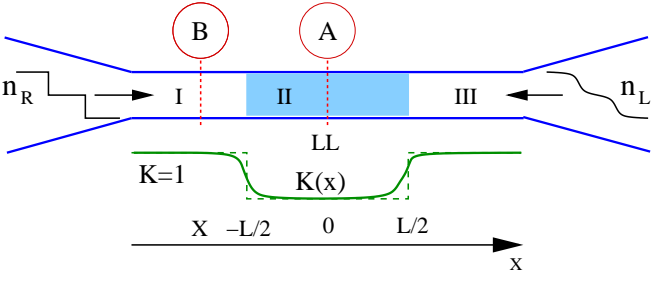}
\caption{Schematic view of noise detection in LL conductor.
The positions of the charge detector inside (A) and outside (B) of
interacting region are indicated. The $x$-dependent LL parameter
$K(x)$ for smooth (solid line) and sharp (dashed line) boundaries is
also shown.}
\label{setup}
\end{figure}

We consider a LL conductor geometry schematically shown in
Fig.~\ref{setup}. Our goal is to solve FCS problem in the presence
of electron interaction. We assume that the spatially dependent
interaction strength $g(x)$ vanishes outside the interval
$[-L/2,L/2]$; this way of modeling leads was introduced in
Refs.~\onlinecite{Maslov,Safi,Ponomarenko} to study the conductance
of LL wire. To describe the FCS of charge transfer, we define a
probability distribution function $p_\tau(n)$   for $n$ electrons to
pass through the cross-section during the time interval $\tau$.   It
is convenient to introduce a generating fuction
$\kappa(\lambda)=\sum_{n=-\infty}^\infty p_\tau(n)e^{in\lambda}$.

Let us first discuss the  non-interacting case. In that case
$\kappa(\lambda)$  has been calculated in
Ref.~\onlinecite{Levitov-noise} by means of Landauer
scattering-state approach. For an ideal quantum wire  (with no
scattering inside the wire)  the generating function of FCS  is
given by
\begin{equation}
\label{generating_function}
\kappa(\lambda)=\Delta_R[\delta_R(t)]\Delta_L[\delta_L(t)]\,.
\end{equation}
Here $\eta=R,L$ labels right-and left-movers and
\begin{equation}
\label{Fredholm_determinant} \Delta_\eta[\delta_\eta(t)]=
\det[1+(e^{-i\delta_\eta}-1)n_\eta]\,
\end{equation}
 is a  Fredholm determinant (of Toeplitz type) of the counting
 operator built of
fermionic distribution function $n_\eta(\epsilon)$  and a
time-dependent scattering phase $\delta_\eta(t)$, with $\epsilon$
and $t$ understood as canonically conjugate variables. In
Eq.~(\ref{generating_function}) and below we assume normalization of
the determinant $\Delta_\eta$ to its value for equilibrium,
zero-temperature distribution. The phase $\delta_\eta(t)$ is given
by
\begin{equation}
\delta_\eta(t)=\lambda \eta w_\tau(t,0)\,,
\end{equation}
where we have defined a window function
$w_\tau(t,\tilde{t})=\theta(\tilde{t}-t)-\theta(\tilde{t}-t-\tau)$.
We use the convention that in formulas $\eta$ is understood as
$\eta=\pm 1$ for right-/left-movers. In the long time limit  the
Fredholm determinant (\ref{Fredholm_determinant})   can be easily
evaluated analytically; in a more general situation it  can be
efficiently studied  numerically  \cite{Nazarov}. Below we
show that for the interacting case, the generating function of the
FCS obeys the form of  Eq.~(\ref{generating_function}). All electron
interaction effects  are  encoded in the time dependent scattering
phases  $\delta_\eta(t)$.

On the microscopic level  the problem  is described by the Keldysh
action $S=S_0[\psi]+S_{\rm ee}[\psi]$, written in terms of fermionic
fields\cite{Kamenev},
\begin{eqnarray*}
\label{TL}
&& S_0[\psi]=i\sum_{\eta}\int_c dt \int dx \psi^\dagger_\eta
\partial_\eta\psi_\eta \,, \nonumber \\&&
S_{\rm ee}[\psi]=-\sum_\eta\int_c dt \int dx g(x) (\rho_{\eta}\rho_{-\eta}+
\rho_{\eta}\rho_{\eta})\,.
\end{eqnarray*}
Here $\rho_\eta=\psi^\dagger_\eta\psi_\eta$ are density fields,
$\partial_{\eta} = \partial_t+\eta v\partial_x$, $v$ is the Fermi
velocity. The non-equilibrium  bosonization approach 
allows us to reformulate  this theory in terms of bosonic (density)
fields. The interacting part of the action, $S_{ee}$, is already
expressed in terms of density modes $\rho_\eta$. Following rotation
in Keldysh space $S_{ee}= \sum_{\eta\eta'}\int dt dx g(x) \rho_\eta
\bar{\rho}_{\eta'}$, while the
 free part of the action, where  information
concerning the state of the non-interacting fermionic system is
encoded,  reads
\begin{eqnarray}&&
\label{action}
S_{0} =\sum_\eta
(-\rho_{\eta} \Pi_\eta^{a^{-1}}\bar{\rho}_{\eta}
-i\ln Z_\eta[\bar{\chi}_\eta]).
\end{eqnarray}
Here we have decomposed the bosonic variables  into classical and
quantum components, $\rho,\bar{\rho}= (\rho_+\pm\rho_-)/\sqrt{2}$,
where the indices $+$ and $-$ refer to the two branches of the
Keldysh contour; $\Pi_\eta^a$ is the advanced component of
polarization operator, and $Z_\eta[\bar{\chi_\eta}]$ is a partition
function of free fermions moving in the field
\begin{equation}
\label{c1}
\bar{\chi}_\eta=\Pi_\eta^{a^{-1}}\bar{\rho}_{\eta}\,,
\qquad
\Pi_\eta^a(\omega,q) = \eta q/2\pi(\eta vq-\omega-i0)\,.
\end{equation}
Expansion of $Z_\eta[\bar{\chi}_\eta]$ in $\bar{\chi}_\eta(t,x)$
generates an infinite series, $i\ln
Z_\eta[\bar{\chi}_\eta]=\sum_n(-1)^{n+1}\bar{\chi}_\eta^n{\cal
  S}_{\eta n}/n$,
governed by irreducible fermionic  density correlation functions,
${\cal S}_{\eta n}(t_1,x_1;\ldots;t_n,x_n) \equiv
\langle\langle\rho_{\eta 1}\rho_2 \dots \rho_{\eta
n}\rangle\rangle$, representing cumulants of quantum noise
\cite{Levitov-noise}.

To find the generating function for the FCS of charge transferred
through a cross-section $x=X$ during a time interval $\tau$, one
needs to calculate the correlation function
\begin{equation}
\label{characteristic_function}
\kappa(\lambda,X)=\langle e^{i\lambda Q(\tau,X)}e^{-i\lambda Q(0,X)}\rangle\, ,
\end{equation}
where  $Q(t,X)=\int_{-\infty}^X (\rho_R(x)+\rho_L(x))dx$ is the
electric charge situated on the left side of the point $X$ at the
time $t$. We note a mathematical similarity  between the problem of
FCS and  that of  tunneling spectroscopy \cite{GGM_2009}. The latter
amounts to evaluation of a single-particle Green function. In both
cases one needs to find a correlation function of exponentials of
bosonic operators that are linear  combinations of the right and
left density fields. A major difference stems from the fact that the
operators $e^{\pm i\lambda Q}$ in
Eq.~(\ref{characteristic_function}) contain a sum of left and right
densities in the exponential, whereas the fermionic operators
$\psi_\eta$, $\psi^\dagger_\eta$ contain only $\rho_\eta$.

We now proceed with calculation of the generating function
(\ref{characteristic_function}); technical details are outlined in the
Supplementary Material \cite{supplementary}. We find that the FCS
generating function has
the form of Eq.~(\ref{generating_function}), as in the case of free
fermions. The time dependent scattering phases $\delta_\eta(t)$ are
expressed through plasmon reflection and transmission coefficients
$r_\eta$, $t_\eta$ at the left (I/II; $\eta=L$) and right (II/III;
$\eta=R$) boundaries;
$r_\eta^2+t_\eta^2=1$. In general, multiple scattering off these
boundaries will give rise to  an  infinite sequence of scattering
phase pulses [cf. Eq.~(\ref{pulses_series})]. If the boundary is
smooth (on the scale of the plasmon wave length), there is no
plasmon reflection, $r_\eta\simeq 0$. In the opposite limit of a
sharp boundary we have $r_\eta = (1-K)/(1+K)$.

\begin{figure}
\includegraphics[width=1\columnwidth,angle=0]{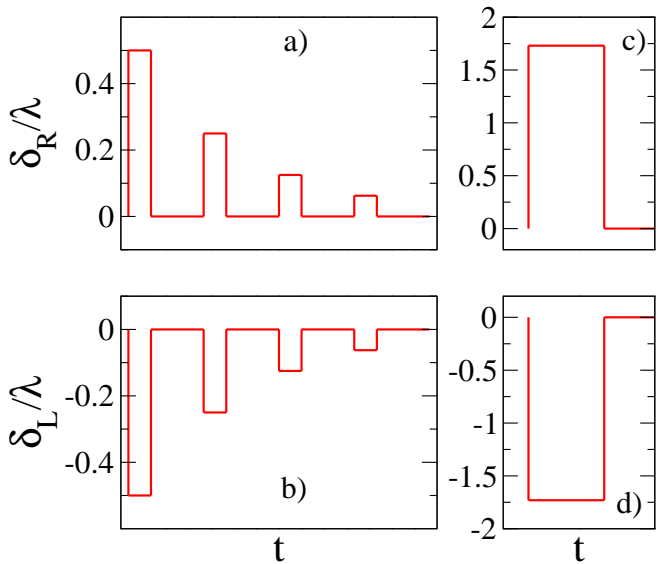}
\caption{Phases $\delta_\eta$ entering
Eq.~(\ref{generating_function}) for the measuring device located at
$X=0$, for sharp (a,b) and adiabatic (c,d) boundaries ($K=1/3$).}
\label{signals}
\end{figure}

In contrast with the  free fermion case, the statistics of current
fluctuations in the LL conductor depend on the position
of the measuring device  
(see Fig.\ref{setup}). Let us first consider the case where the
current fluctuations are measured  in the middle of an  interacting
region ($X=0$, position A in Fig.\ref{setup}). 
We thus find \cite{supplementary} that time dependent scattering
phases consist of  a sequence of 
pulses
\begin{equation}
\label{pulses_series}
\delta_\eta(t)=\sum_{n=0}^\infty\delta_{\eta,n}w_\tau(t,t_n)\,
\end{equation}
with  partial phase shifts
\begin{eqnarray}&&
\label{phase_center}
\delta_{\eta,2n}=\eta\lambda t_{-\eta}\sqrt{K} r_\eta^nr_{-\eta}^n
\equiv \eta\lambda e^*_{\eta,2n} \,, \\&&
\delta_{\eta,2n+1}=\eta\lambda t_{-\eta}\sqrt{K}
r_\eta^{n+1}r_{-\eta}^n \equiv \eta\lambda e^*_{\eta,2n+1} \,.
\end{eqnarray}
The beginning of the $n$-th pulse occurs at time
$t_n=(n+1/2-1/2K)L/u$.
 The sequence of pulses  $\delta_\eta(t)$  is shown
in Fig.~\ref{signals} for the  limiting cases of sharp and  smooth
boundaries between the interacting and non-interacting regions.

In the limit of very low frequencies ($\tau \gg L/u$)  all  pulses
(\ref{pulses_series})  overlap and the scattering phase can be
approximated by $\delta_\eta(t)\simeq
w_\tau(t,0)\sum_{n=0}^\infty\delta_{\eta,n}$. In this limit of long
wavelength the
interacting/non-interacting  boundaries appear to be  sharp. This
yields $\sum_{n=0}^\infty\delta_{\eta,n}=\eta\lambda$. As expected
\cite{Maslov,Safi,Ponomarenko} the effects of interaction in this
limit disappear and one is back to the FCS of free
fermions\cite{Levitov-noise}.

In the opposite case ($\tau \ll L/u$ )  the interference between
different   plasmon pulses  may be neglected
and the result splits into a product
\begin{equation}
\Delta_\eta[\delta_\eta(t)]\simeq \prod_{n=0}^\infty
\Delta_{\eta\tau}(\delta_{\eta,n})\,,\qquad \delta_{\eta,n} \equiv
\eta\lambda e^*_{\eta,n}\,,
\label{determinant-product}
\end{equation}
with each factor representing  a contribution of a single phase
pulse $\delta_{\eta,n}(t)=\delta_{\eta,n}w_\tau(t,0)$. Remarkably,
plasmon scattering gives rise to  charge fractionalization, which
splits the scattering phase into an infinite series of pulses. On a
technical level, the Fredholm determinant of the counting operator
is now replaced by an infinite product  of determinants, each
calculated with a corresponding scattering phase. The FCS of the LL
then becomes a superposition of FCS of non-interacting electrons
with fractional charges $e^*_{\eta,n}$. For the case of smooth
boundaries we get only one fractional charge, $e^*_{\eta,0}=
\sqrt{K}$. In the opposite limit of sharp boundaries, we obtain the
sequence of fractional charges of the form $e^*_{\eta,n}=
2K(1-K)^n/(1+K)^{n+1}$.

Focussing on the regime where the counting interval $\tau$ is long
compared to the inverse energy scale of the distributions
$n_\eta(\epsilon)$, we express $\Delta_{\eta\tau}(\delta)$ in the
form
\begin{equation}
\ln \Delta_{\eta\tau}(\delta) = \tau\! \int\!\! {d\epsilon \over 2\pi} \{\ln
  [1+(e^{-i\delta}-1)n_\eta(\epsilon)] + i\delta\theta(-\epsilon)\}.
\label{determinant-semiclassics}
\end{equation}
Substituting (\ref{determinant-semiclassics}) in
Eqs.~(\ref{generating_function}), (\ref{determinant-product}) and
expanding in $\delta$, one can find explicitly all moments of noise.

Consider now the FCS   outside the interacting region, e.g. at $X
<-L/2$ (B in Fig.\ref{setup}). The analysis follows closely the one
outlined above, hence    we present only the final result. The right
and left scattering phases are
\begin{eqnarray}
\delta_R(t)&=&\lambda\bigg[w_\tau\left(t,{X\over v}\right)-
r_Lw_\tau\left(t,\frac{L+X}{v}\right) \nonumber
\\ \nonumber
&+& r_Rt_L^2\sum_{n=0}^\infty
r_L^nr_R^nw_\tau\left(t,\frac{L+X}{v}+{2(n+1)L\over u}
\right)\bigg]\,,\nonumber \\
\delta_L(t)&=&-\lambda t_Lt_R \sum_{n=0}^\infty r_L^nr_R^n
w_\tau\left(t,{2nL\over u}-{X\over v}+\frac{v-u}{vu}L\right)\,.
\nonumber
\end{eqnarray}
Current fluctuations measured in the non-interacting region at
finite frequency ($\tau \ll L/u$) differ from those at the
interacting part of LL (\ref{phase_center}). Here too  the FCS takes
the  form of superposition of non-interacting FCS's,  but with a new
set of fractional charges:
\begin{eqnarray}
&& e^*_{R,0}=1\,, \ \ e^*_{R,1}= - r_L\,,\ \  e^*_{R,n+2}=
t_L^2r_L^nr_R^{n+1}\,, \nonumber \\
&& e^*_{L,n}=t_Lt_Rr_L^nr_R^n\qquad(n=0,1,\ldots).
\label{charges-nonint}
\end{eqnarray}
In the universal  ultra-low-frequency  limit ($\tau \gg L/u$) the
effects of interaction again disappear, and the result for the  FCS
of free fermions is recovered.

To conclude we provide a brief overview of the physics discussed
here. We have calculated the FCS in a LL conductor. The result is
cast in terms of a Fredholm determinant of the counting operator.
The boundaries between the interacting and non-interacting regions
of the wire give rise to (in general, multiple) plasmon
scattering. This generates 
plasmon wavelets corresponding to fractionally charged ``particles''. 
As a result, for a counting time interval
$\tau$ small compared with the plasmon time-of-flight through
the interacting region, $L/u$, the FCS is a superposition of counting
statistics of  non-interacting particles with fractional charges
$e^*_{\eta,n}$. Let us stress that the FCS contains more complete
information about the system than the second moment (noise). As an
example, for a symmetric system ($r_L=r_R$) the noise acquires the
factor $K$ due to interaction. By itself, this is not sufficient to
make a conclusion about the value of fractional charges and the
character of correlations between them. 

The fractionalization process that manifests itself in the FCS
caculated above is solely due to the interacting/non-interacting
boundary scattering \cite{Safi}. Fractionalization due to tunneling
into a LL \cite{lehur} does not show up here: technically this is
since  the operator $Q$ in Eq.~(\ref{characteristic_function}) is
determined by a sum of $\rho_L$ and $\rho_R$. 

How is the notion of charge fractionalization compatible with charge
quantization? The above analysis was based on the bosonization
approach with the density $\rho_\eta$ slowly varying on the scale of
the inverse Fermi wave vector $k_F^{-1}$. This corresponds to the
situation where the measurement procedure is smooth on the scale
$k_F^{-1}$. In this situation the above results are valid for any
counting field $\lambda$: our detector may count a fractional charge
residing in some volume of space. In the opposite limit, where the
spatial resolution of our measurement is sharp on the scale
$k_F^{-1}$, the measured charge should be integer, i.e. the FCS
should satisfy $\kappa(\lambda)=\kappa(\lambda+2\pi)$. In the
bosonization formalism,  charge quantization emerges  by taking into
account fast oscillatory contributions to the density $\rho_\eta$
\cite{Haldane}. This will not affect the moments of FCS (which are
obtained as derivatives of $\kappa(\lambda)$ at $\lambda=0$) and, more
generally, the form of $\kappa(\lambda)$ in the range $[-\pi,\pi]$.
Beyond this interval, the FCS will be  continued periodically,
in agreement with the charge quantization requirement.

This work was supported by GIF, DFG Center for Functional Nanostructures,
Einstein Minerva Center, US-Israel BSF, ISF, Minerva Foundation, DFG
SPP 1285, EU GEOMDISS, and Rosnauka 02.740.11.5072.

\newpage

\appendix
\section*{ONLINE SUPPLEMENTARY MATERIAL}

\renewcommand{\theequation}{S\arabic{equation}}
\setcounter{equation}{0}

Here we present details of the calculation of the generating function
(\ref{characteristic_function}). Despite the fact that the bosonic
action (\ref{action}) contains terms of all orders, evaluation of
the generating function  can be performed exactly. The point is that
the action (\ref{action}) is linear with respect to the classical
component $\rho_\eta$ of the density field. Thus, integration with
respect to $\rho_\eta$  can   easily be   performed. It  gives rise
to a $\delta$-function condition for the quantum component
$\bar{\rho}_\eta$ of the density,
\begin{eqnarray}&&
\label{a7}
\partial_t\bar{\rho}_{R}+\partial_{x}\left
[(v+{g\over 2\pi})\bar{\rho}_{R}+\frac{g}{2\pi}\bar{\rho}_{L}\right]
=\frac{\lambda}{2\pi}j\,,
\nonumber \\&&
\partial_t\bar{\rho}_{L}-\partial_{x}\left[(v+\frac{g}{2\pi})\bar{\rho}_{L}+
\frac{g}{2\pi}
\bar{\rho}_{R}\right]=-\frac{\lambda}{2\pi}j\,.
\end{eqnarray}
Here $j_{\tau}(x,t)$ is a source term,
\begin{equation}
j_{\tau}(x,t)= \delta(x-X)[\delta(t-\tau)-\delta(t)]/\sqrt{2}\,.
\end{equation}
It remains to solve   Eqs.~(\ref{a7}) for $\bar{\rho}_\eta$, to calculate
$\bar{\chi}_\eta$ according to Eq.~(\ref{c1}) and then to substitute it
into the second term of the bosonic action (\ref{action}). This yields
\begin{equation}
\label{a2}
Z_\eta[\bar{\chi}_\eta] =
\Delta_\eta[\delta_\eta(t)].
\end{equation}
We thus find that the generating function of  the FCS problem has
the form of Eq.~(\ref{generating_function}), as in the case of free
fermions. The time dependent scattering phases $\delta_\eta(t)$ are
given by the integral of $\bar{\chi}_\eta$ along a free-electron
trajectory,
\begin{eqnarray}
\label{d1a}
\delta_\eta(t) &=&\sqrt{2} \int_{-\infty}^\infty
d\tilde{t}\: \bar{\chi}_\eta(\tilde{t}-t,\eta v\tilde{t})\nonumber \\
&=&-2\pi\sqrt{2}\:
\eta\lim_{\tilde{t}\rightarrow-\infty}\int_0^{\eta v(\tilde{t}+t)}d\tilde{x}\:
\bar{\rho}_\eta(\tilde{x},\tilde{t})\,.
\end{eqnarray}
The second line of Eq.~(\ref{d1a})
expresses $\delta_\eta(t)$ through the
asymptotic behavior of $\bar{\rho}_\eta(x,t)$ in the non-interacting
parts of the wire (regions I and III in Fig.\ref{setup}).

To find the scattering  phases, we now solve the set of linear
differential equations (\ref{a7}). These equations can be
conveniently combined into a single differential equation for the
current operator $\bar{J}=v(\bar{\rho}_R-\bar{\rho}_L)$
\begin{equation}
\label{J_equation}
(\omega^2+\partial_x u^2(x)\partial_x)\bar{J}(\omega,x)=0\,, \,\,\, x \neq X\,,
\end{equation}
where $u(x)= v/K(x)$ is the plasmon velocity and $K(x) = [1+g(x)/\pi
v]^{-1/2}$ is the LL parameter. The latter is equal to 1 in the leads
and takes a constant value $K$ in the interacting region; the
crossover between these values depends on the profile of the
interaction.
Inside the interacting region $\bar{J}(\omega,x)$  is given by
a superposition of the right and left moving plasmon waves:
\begin{equation}
\label{J_waves}
\bar{J}(\omega,x)
=u^{-1/2}(a e^{i\kappa x}+ b e^{-i\kappa x})\,, \qquad \kappa=\omega/u\,.
\end{equation}
In the non-interacting regions, an analogous form holds, with the
replacement $u \to v$.
%
The change of interaction strength constitutes a source of plasmon scattering
and can be described by plasmons scattering matrices,
\begin{equation}
\label{s-matrices}
S_L = \left(
\begin{array}{cc}
t_L & - r_L \\
r_L & t_L
\end{array}
\right), \qquad
S_R = \left(
\begin{array}{cc}
t_R &  r_R \\
- r_R & t_R
\end{array}
\right).
\end{equation}
The matrices $S_{L,R}$ relate the amplitudes $a,b$ at the left
(I/II) and right (II/III) boundaries, respectively;
$r_\eta^2+t_\eta^2=1$. In general, multiple scattering off these
boundaries will give rise to  an  infinite sequence of scattering
phase pulses (cf. Eq. \ref{pulses_series}). If the boundary is
smooth (on the scale of the plasmon wave length), there is no
plasmon reflection, $r_\eta\simeq 0$. In the opposite limit of a
sharp boundary we have $r_\eta = (1-K)/(1+K)$. 

Using Eqs.~(\ref{d1a}),
(\ref{J_equation}), (\ref{J_waves}), and (\ref{s-matrices}), we obtain
the phases $\delta_\eta(t)$ as given in the main text of the paper.

\end{document}